\documentstyle[aps,epsf,preprint,prl]{revtex}
\def\myfigure#1#2{{\leftskip=0.000753\textwidth \rightskip\leftskip\small
\begin{figure}\baselineskip=24pt plus 2pt minus 1pt
\centerline{#1}\nobreak\smallskip\nobreak #2\end{figure}}}
\global\firstfigfalse 
\draft

\newcommand{\bea}{\begin{eqnarray}}
\newcommand{\eea}{\end{eqnarray}}
\newcommand{\be}{\begin{equation}}
\newcommand{\ee}{\end{equation}}
\newcommand{\nn}{\nonumber}

\newcommand{\beq}{\begin{equation}}
\newcommand{\eeq}{\end{equation}}
\newcommand{\bdm}{\begin{displaymath}}
\newcommand{\edm}{\end{displaymath}}       
\begin{document}                               
\title{Dynamics of folding in Semiflexible filaments}
\author{P. Ranjith \footnote{Email:ranjith@physics.iitm.ac.in}
 and P.B. Sunil Kumar\footnote{Email:sunil@physics.iitm.ac.in}}
\address{Department of Physics,
Indian Institute of Technology, Madras\\
Chennai 600 036,\\
India}
\date{\today}
\pagebreak
\maketitle

\begin{abstract}
We investigate the dynamics of a single semiflexible filament, under the action of a
compressing force, using numerical simulations and  scaling arguments.
The force is applied along the end to end vector at one 
extremity of the filament, while the other end is held fixed.  
We find that, unlike in elastic rods the filament folds asymmetrically with a
folding length  which depends only on the  bending stiffness 
$\kappa$ and the applied force $f$. It is shown that this behavior 
can be attributed to the exponentially falling  tension profile in the filament.
While the folding time $\tau_0$ depends on the initial configuration, 
the distance moved by the terminal point of the filament and the length of the fold 
scales as $\tau^{1/2}$ at time $\tau>>\tau_0$ and is independent of the initial 
configuration.
\end{abstract}


Recent developments in micro 
manipulation techniques using optical tweezers, micro pipette and high 
resolution optical 
imaging have made the study of  single filament dynamics experimentally feasible
~\cite{strick,ashkin,Dogte}. Today it is possible to apply forces of the order of 
pico Newton with precision.  This has spurred lot of activity in  the
study of single bio-polymers. These techniques have
 been used extensively in the study of stretching elasticity and force extension
 relation  of bio-polymers ~\cite{strick,poptic,doptic}. Some of these techniques
can also be applied to study the effect of compressing forces on single polymer 
filaments ~\cite{poptic,doptic}.

In a physical description of their statistical properties, the chemical details
of polymers can be summarized using a set of parameters.
These parameters characterize the individual features of each polymer species.
Two of the most important parameters are persistence length and bending 
stiffness.
Persistence length, $l_{p}$, is the distance along the contour of the 
filament over which the tangent-tangent correlation is destroyed by thermal 
fluctuations. Semiflexible polymers have contour length comparable to its 
persistence length.  This means that unlike in the case of  flexible filaments or 
rigid rods the dynamical properties of semiflexible filaments  are determined by both
conformational entropy and bending stiffness. 
Thermal fluctuations also play an important role in determining the 
viscoelastic behavior 
of semidilute solutions of semiflexible filaments ~\cite{Maggs,Frey}.

From the biological point of view, a study of the mechanical and statistical mechanical 
properties of semiflexible filaments are important to understand the  dynamical 
behavior of cytoskeletal network.  Cytoskeleton, which controls the shape recovery 
and dynamic cellular reorganization of living cells is a complex network of 
polymers ~\cite{Alberts}.  At the length scale of a cell most of the polymers 
which build up this network are  semiflexible ~\cite{Janmey}. Microtubules, 
$l_p=5.2 mm$ ~\cite{Gittes1}, and Actin filaments, $l_p=17 \mu m$ ~\cite{Ott} 
are  some of the examples.

For length scales much smaller than the persistence length the filaments are 
like rigid rods. Under the action of longitudinal compressing forces these 
filaments exhibit the classical Euler buckling ~\cite{landau}. 
The dynamics of this 
buckling has been studied recently ~\cite{golu}. In the case of bio-polymers like 
microtubules, such buckling 
is also observed as a result of the 
force generated by polymerization~\cite{Dogte} and by the movement, on them, of kinesin motors 
~\cite{Gittes2}.  On the other hand 
in flexible polymers with persistence length much smaller than the contour length, 
the compression forces applied at the extremities do not propagate along the polymer.
This naturally raises the question, what happens when we apply a longitudinal 
compressing force on a semiflexible filament?   Here we investigate the dynamics of a single semiflexible 
filament under the action of an external compressing  force.  

In this paper we look at the folding of a semiflexible filament under the action of 
a compressing force acting along the end to end vector.
Previous calculations of folding in semiflexible 
polymers  assumes uniform tension in the filament ~\cite{odijk,golu}.
We show that, at finite temperature,  the  tension profile in a semiflexible filament in equilibrium decreases
exponentially along the contour. This tension profile 
is shown to have  non trivial effects on the folding dynamics of filaments.
The filament is described by the position
vector 
${\bf r}(s)$, parameterized by the contour length $s$ of the filament, where $0 \leq s \leq L$.             
 Variety of boundary conditions are possible. Throughout this paper we will
 discuss only the case wherein the curvature of the filament at the two extreme
 ends are zero (hinged boundary). This implies free rotation of the tangent vector 
at the ends.
The $s=0$ tip of the filament is allowed to move only along the direction of 
the end to end vector which is taken to be along the $x-$axis. The position of
the $s=L$ tip, ${\bf r}(L)$, is fixed. 
An external force ${\bf f}$ is applied at the $s=0$ tip along the $x-$direction.
  A detailed discussion of other possible 
boundary conditions will be published elsewhere ~\cite{ranjith}. 
The elastic
 coefficients for the transverse and longitudinal fluctuations of semiflexible
 filaments are not the same. This implies that, for the above  boundary
 conditions,  a compressing force could lead to either folding or buckling.
 However, the transverse spring coefficient in 
the stiff limit is proportional to stiffness  $\kappa$ and the longitudinal spring coefficient 
proportional to $\kappa^2/T$ ~\cite{frey}. Thus an applied force on a filament with 
hinged ends will always lead to folding. 

Actin filaments exhibit a persistence length of about $17 \mu m$ at
 room temperature ~\cite{Ott}. Critical force for buckling is then of the 
order of a few pico Newton.  In a laser tweezers experiment  we can 
apply forces of the order of pico Newton with precision. Typically, in 
such experiments the filament is held by a bead of size $\sim 0.5 \mu m$ 
to a precision of $\sim \pm 50 nm$ ~\cite{Wang}.
In view of these, it should be possible to observe  folding dynamics 
in laser tweezers experiments using actin filaments of length
 $< 10 \mu m$ or microtubules of length $> 100 \mu m$ .\\

In our numerical simulations, semiflexible filament are modeled as $N$ rods 
of equal length hinged together linearly. 
Position of each hinge is defined by the Cartesian 
coordinates ${\bf r}_{i}=(x_{i}, y_{i})$.  
The total energy of the filament, with the compressing force {$\bf f$} 
acting at ${\bf r}_1$, is  ~\cite{bensimon}, 

\beq
F = \frac{\kappa}{2} \sum_{i=1}^{N-2} (1- {\bf t}_{i}\cdot{\bf t}_{i+1})+\sum_{i=1}^{N-1}g_{i}{\bf t}_{i}^2 -x_{1} {\bf \hat e_{x}} \cdot { \bf f},\label{kratky
e}
\eeq

where ${\bf t_{i}}$ is the unit vector pointing from $i$ to 
$i+1$. 
The first term is the contribution from the curvature of the polymer, $\kappa$ being the elastic constant,  and the second term 
is to enforce  the local inextensibility condition, ${\bf t_{i}}^2 =1 $,  of the rods.
We work in the over damped limit where the dynamics of the
 filament is described by  the Langevin equations:\\
\bea
\Gamma^{-1}\frac{\partial x_{i}} {\partial t}&=& - \frac{\partial F} {\partial x_{i}} + \eta_{x_{i}} (t) \nonumber \\
\Gamma^{-1}\frac{\partial y_{i}} {\partial t}&=& - \frac{\partial F} {\partial y_{i}} + \eta_{y_i} (t).
\label{dic_lang}
\eea
Here the effect of the solvent enters only through the inverse 
friction coefficient $\Gamma$.
Equations~\ref{dic_lang} are solved along with the constraint ${\bf t}_i^2=1$.
The noise $\eta$ obeys the fluctuation dissipation theorem which in the case of 
filaments with equal monomer mass leads to~\cite{hinge} 
\bdm
\left<\eta_{x_i}(t) \eta_{y_j}(t')\right> = 2\Gamma^{-1} \delta_{xy} \delta_{ij} \delta(t-t').
\edm

Integrating equations ~\ref{dic_lang} we get the increments $\Delta x_i$ and 
$\Delta y_i$, which are functions of the unknown Lagrange multipliers $g_i$.
Increment in time is represented by the dimensionless variable  
$ \Delta \tau = \Gamma \Delta t $ .
The coordinates of the vertices at time $\tau+\Delta \tau$ 
obtained from the above equations
have to satisfy the constraint ${\bf t}_i^2(\tau +\Delta \tau)=1$.
These $N-1$ simultaneous nonlinear equations are solved to determine
the Lagrange multipliers $g_i$.  We choose $\Delta x_N=0$,
$\Delta y_N=0$ and $\Delta y_0=0$ at all times.
 There is no constraint
on the angle between the filament and the force direction at both ends.

We have carried out extensive numerical simulation of the above model using a chain
with length $N=150$ for different values of $\kappa$ and $f$. One run in the 
simulation correspond to $\Delta \tau \,=\,1$. All results
are averaged over 10 initial configurations.
The initial conformation of the filament $(\tau=0)$ is chosen from an equilibrium 
distribution which satisfy the boundary condition $y_1=0 $ and $y_{N}=0$ ~\cite{udo}.
The snapshots from our simulation for a constant applied force $f$ is shown in
figure~$1$. 
We observe that the filament first develop a fold at a distance $l_{f}$ from the end at which 
the force is applied. $l_f$ is the position of the first maximum in curvature along 
the contour when the free end is perpendicular to the end to end vector. For a given $\kappa$ and force $f$, this "folding length"$(l_{f})$ 
is independent of the initial configuration and  is found to 
depend on bending stiffness as 
$l_f \sim \sqrt{\kappa}$. This is shown in figure~$2$.  

\myfigure{\epsfysize2.0in\epsfbox{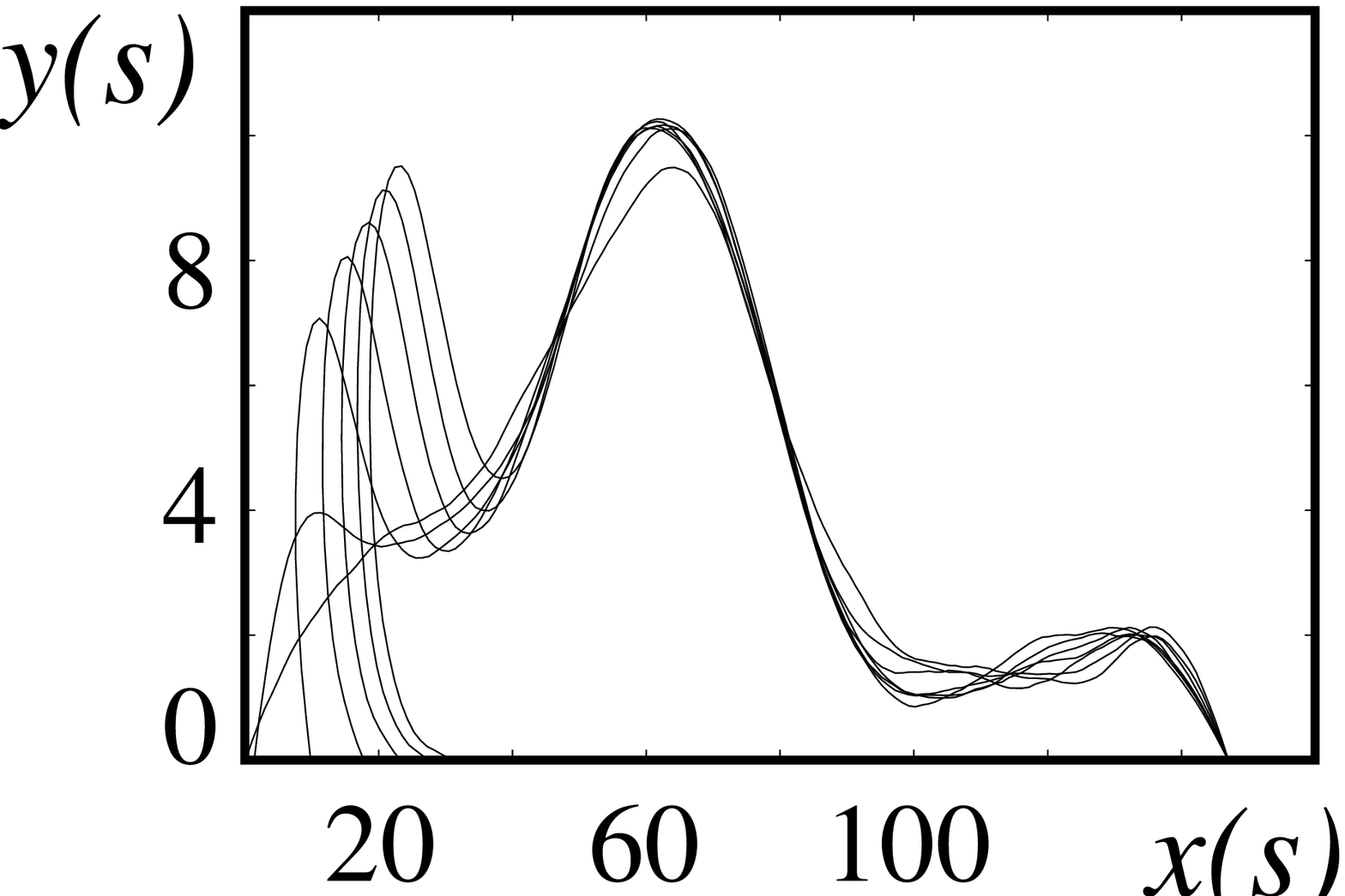}}{\vskip-0.00in}
FIG. $1$ The snapshots of conformations of a semiflexible filament  with $\kappa=500$
and $N=150$, taken at equal intervals of time $\Delta \tau=0.2$ million for an applied force $f=30$.  \\

\myfigure{\epsfysize2.0in\epsfbox{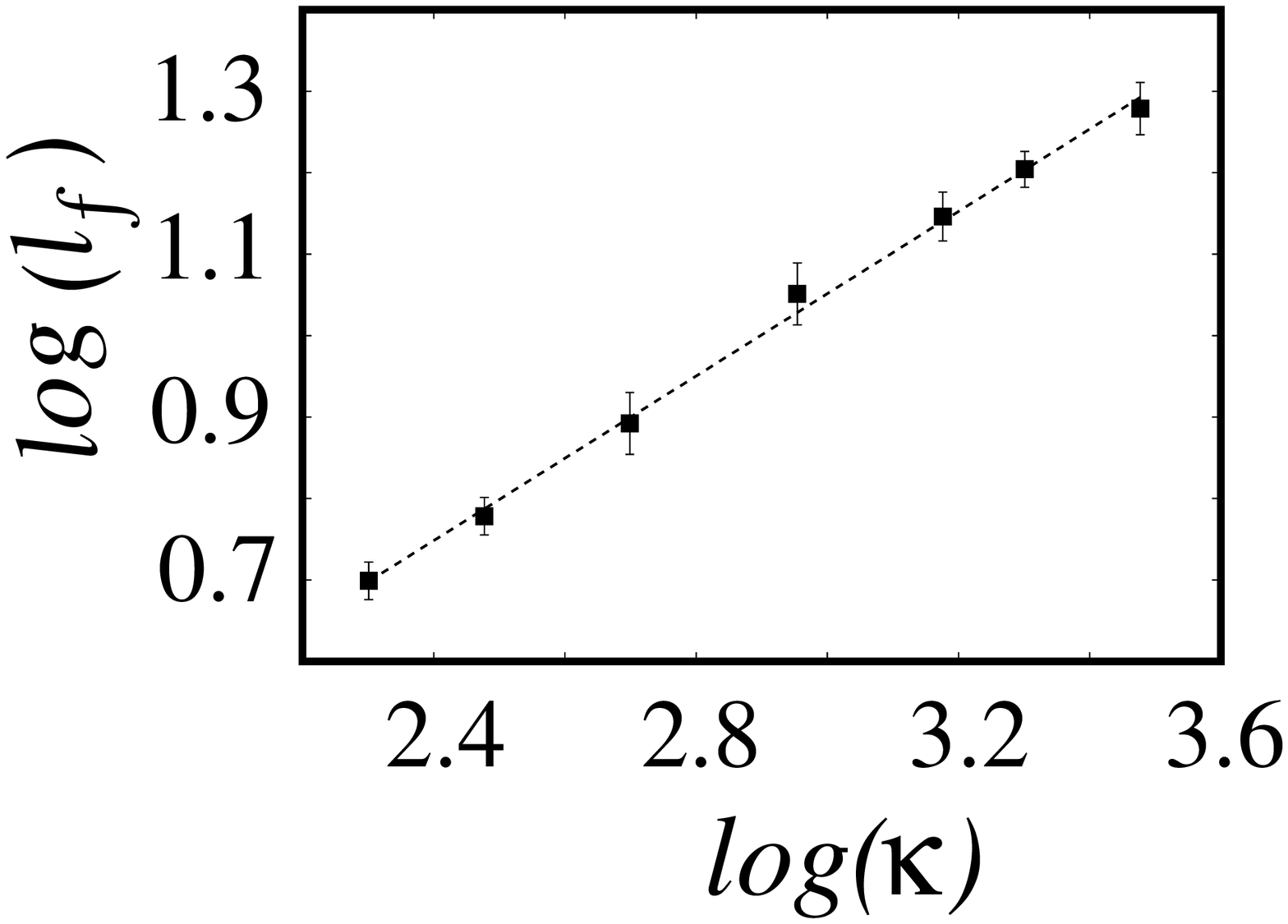}}{\vskip-0.0in}
FIG.$2$ ~ A plot of  $\log_{10}\kappa$  vs $\log_{10}l_{f}$. The filled squares
are from the simulation. The dotted line is a fit which shows 
~ $l_f \sim \sqrt{\kappa}$.\\

A better understanding of the dependence of folding length on $\kappa$ is obtained 
from the continuum equations for the worm like chain model~\cite{Doi,udo} .
The equation of motion in the over damped limit for the boundary conditions, ${\bf r}(L)=$a constant
and $y(0)=0$, is then 
\bea
\frac{1}{\Gamma} \frac{\partial {\bf r}}{\partial t}&=&-  \frac{\delta F}{\delta {\bf r}}=-l_p \partial^4_{s} {\bf r}+ 2\partial_{s} (g \dot{\bf r}) \nn \\
&&+ \left(f+ 2g(s)\dot {\bf r}(s)
-l_p \partial^3_{s} {\bf r}\right) \delta(s) + \eta(s,t),
\label{langc}
\eea 
here dots denote the derivatives with respect to arc length. 
$\eta(s,t)$ is the thermal noise and $\Gamma$ is a kinetic coefficient.
Using  the constraint ${\bf \dot r}(s)^2=1$,  and equation~\ref{langc} we get,
\beq
\frac{1}{2 \Gamma} \frac{\partial {\bf \dot r}^2}{\partial t}= 2\ddot g-2g \ddot r^2 -
l_p \dot {\bf r} \cdot \partial _{s}^5 {\bf r}=0.
\label{geqc}
\eeq

Comparing the curvature and force terms in equation~\ref{langc}, 
we get a length which scales as $l \sim \sqrt(\kappa/f)$. 
Similar length scale is also obtained from the constraint equation. 

Let us try to get some insight into the tension profile along the filament. Dropping higher order 
derivative terms from the constraint  equation~\ref{geqc} leads to
\be
\frac{1}{2 \Gamma} \frac{\partial {\bf \dot r}^2}{\partial t}= \ddot g^f-g^f \ddot r^2=0. \label{geqn1}
\ee                    

We use $g^f$ to distinguish this tension from that obtained with $\kappa$. Numerical solution of the above equation  for time $t=0$ is
shown in figure~$3$. Here we have replaced $\ddot{\bf{r}}(s)^2 $ by its approximate  ensemble
averaged value

\be
\left< \ddot{r}(s) ^2 \right> \,=\,\left( \frac{1}{l_p L}\right) \frac{ \sum_n \sin^2 \left( \frac{n \pi s}{L} \right) }{1-\frac{L}{\kappa \pi} \sum_n \frac{1}{n^2} \sin ^2 \left(\frac{n \pi s}{L}\right) } \label{avgcurv}
\ee
\myfigure{\epsfysize2.0in\epsfbox{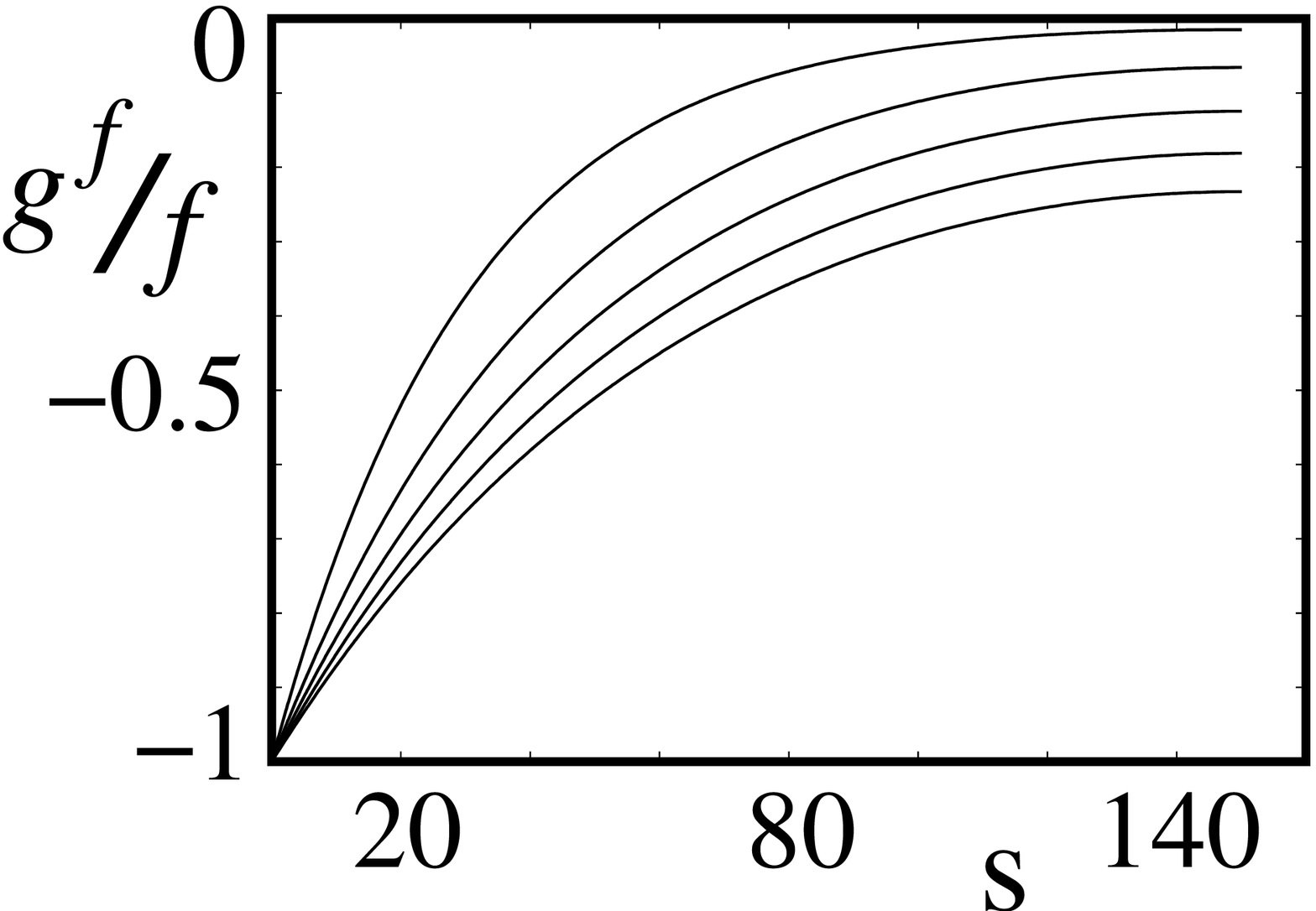}}{\vskip-0.2in}
{FIG.$3$~$g(s)$ obtained from equation~\ref{geqn1} for $\kappa=500,1000,1500,2000,2500$ and 
n=150.}\\

We see that the resulting tension decays exponentially. We find the length scale 
to vary as  $\sqrt \kappa $. 
The tension profile is found to depend on the number of modes taken in  
 equation~\ref{avgcurv} and is exponential only in the limit when the number of 
modes are large. The lower cut off in length being set by the segment length.

We now compare this tension profile  with that obtained from the simulations.
In the simulations, the tension $g_i$ has contributions from three sources,
namely, curvature, noise and  the applied force. 
Like in the case of continuum equations we can calculate a tension $g^f_i$ by 
setting the noise and curvature terms to zero.
Since the equations of motion are nonlinear  one may doubt 
the validity of a comparison between the tension $g^f_i$ obtained from this 
calculation with $g^f(s)$ obtained from the continuum equations. To dispel 
this doubt we linearize the 
equations for the tension profile by neglecting terms of the order  
$(\Delta x_i)^2$ and $(\Delta y_i)^2$, in equation ${\bf t}_i^2=1$. The results
obtained from such a linear calculation agree well with the full non-linear 
results (data not shown).  However, this linearization 
 also  means  deviation from the constraint ${\bf t}_i^2=1$.  Even for small $\Delta \tau$ these errors in  
segment length can become very large in a short time. Therefore we have to 
use the full non-linear equations for updating  the filament 
configuration. 

In figure~$4$ we show this tension profile $g^f_i$ as a function of the 
vertex index $i$. The tension $g^f_i$ goes to zero 
exponentially as predicted earlier. The distance $\xi$ over which the 
tension falls to half its maximum value is found to increase as $\sqrt \kappa$ for
a fixed value of $f$.

\myfigure{\epsfysize2.5in\epsfbox{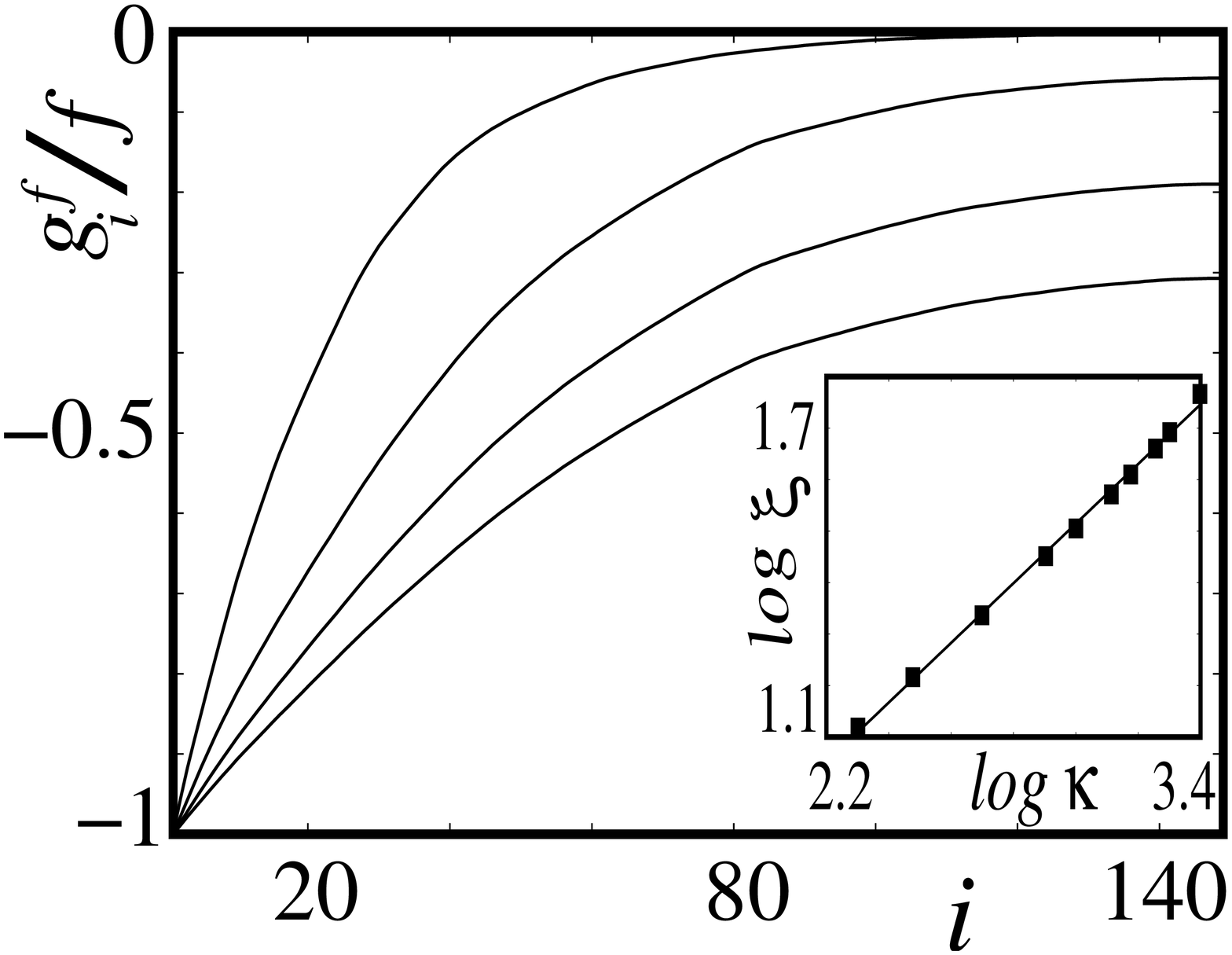}}{\vskip-0.2in}{
FIG.4 $g^f/f$ as a function of the contour length $i$ from the point at which 
force is applied, for $\kappa=300,1000,2000,3000$. Inset: $\log(\kappa)$ is plotted against 
the $\log \xi$, showing $\xi \sim \kappa ^{1/2}$  }\\

We now look at how the buckling length $l_f$ depends on the applied force $f$.
We have seen in the previous paragraph that the tension in a semiflexible filament 
falls exponentially along its contour. This exponential tension profile differentiates between 
semiflexible filaments and stiff elastic rods wherein the tension profile is 
uniform. 
Thus it is instructive to extend the stability calculations for
a rod with uniform tension ~\cite{landau} to the present case of
exponential tension 
profile of the form $\exp(-\alpha s/\sqrt{\kappa})+1$.
The equation of equilibrium is,

\beq
\kappa \frac{\partial^{4}y}{\partial s^4}+\frac{f}{2} \frac {\partial}{\partial s} \left[ \frac{\partial y}
{\partial s} \left[ \exp(-\alpha s/\sqrt{\kappa})+1 \right] \right]  =0.
\label{steq}
\eeq

These equations  are solved numerically with the boundary conditions 
$\ddot{y}|_{s=L}=0$, $\ddot{y}|_{s=0}=0$, $y(0)=0$, $y(L)=0$.
The lowest energy solutions of this equation show that the value  $s=s_{max}$, of the first maximum  in curvature, 
shifts to smaller values of $s$ with increasing $f$.  This is indeed what 
we see in our simulations of the semiflexible filament.

\myfigure{\epsfysize2.0in\epsfbox{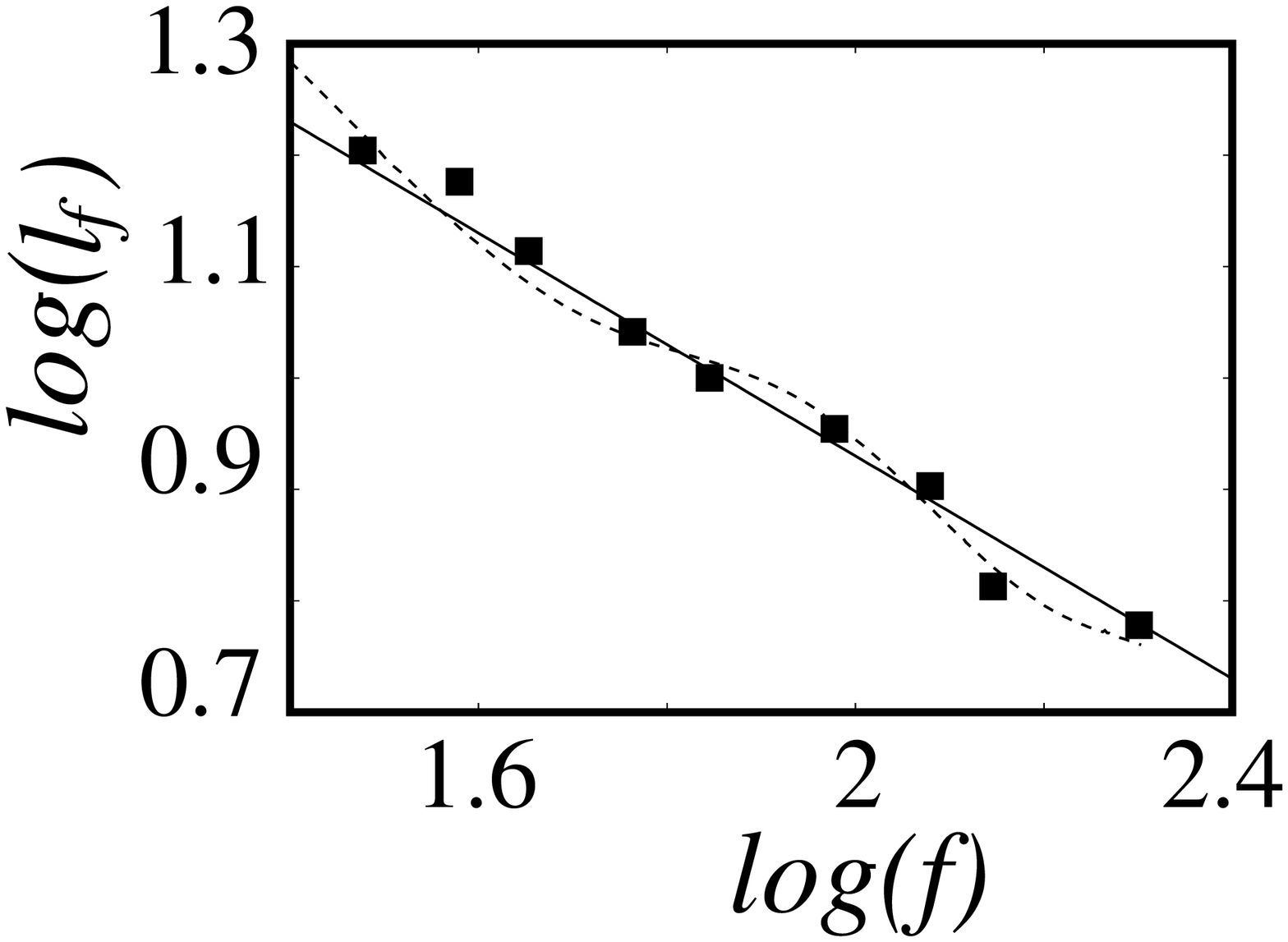 
}}{\vskip-0.2in}{  FIG.$5$  Folding length ($l_f$) as a function of the
applied force($f$); The dotted line represents the buckling point obtained
from the numerical solution of equation~\ref{steq} with $\alpha=10$. 
The straight line corresponds to $l_f \sim \frac{1}{\sqrt{f}}$}\\

In figure~$5$ we compare the buckling length $l_f$  obtained from the simulation 
with $s_{max}$ from the solution of  equation ~\ref{steq}. $\alpha$ is the 
only adjustable parameter used to get the agreement between the two results.

We now come to the dynamical behavior of the fold. The length of the segment 
before the fold, the end to end distance and tension $g^f_i$ all depend on time. As can be seen 
from the configuration 
snap shots in figure~$1$, once the filament has folded the applied force 
acts to {\it pull} the filament. This implies a change in sign of tension 
$g^f_i$.  At $\tau=0$ we have  
$g^f_1=-f$ . As the filament folds $g^f_1$ changes 
sign saturating to $g^f_1=f$ . The time $\tau_0$ at which 
$g^f_1(\tau_0)=0$ can be designated as the folding time.
We find this value of the folding time to depend on the initial configuration 
of the filament. 
The end of the filament at which 
the force is applied, moves  by a distance $x_1(\tau)$ at time $\tau$. 
This is depicted in figure~$6$ along with the length of the fold.

For 
$\tau>>\tau_0$  all the distances scale as $\tau^{1/2}$. This shows that in this regime 
the problem is similar to that of stretching a tense string 
~\cite{udo}. For large $f$  we can then neglect the contribution from 
the curvature energy and noise. The dynamics is then determined by the straight segment 
up to the fold moving through the fluid. This results in a linear tension profile along the
 filament ~\cite{udo}, with the tension  falling to zero approximately  at the fold.
At $s=0, g^f(s=0)=f$, by neglecting the curvature and noise terms on the right hand side 
of the equation ~\ref{langc}, using  
dimensional analysis, we get,  $x(s=0,\tau)\sim (\tau)^{1/2}$.  This is shown in figure~$6$.

\myfigure{\epsfysize2.5in\epsfbox{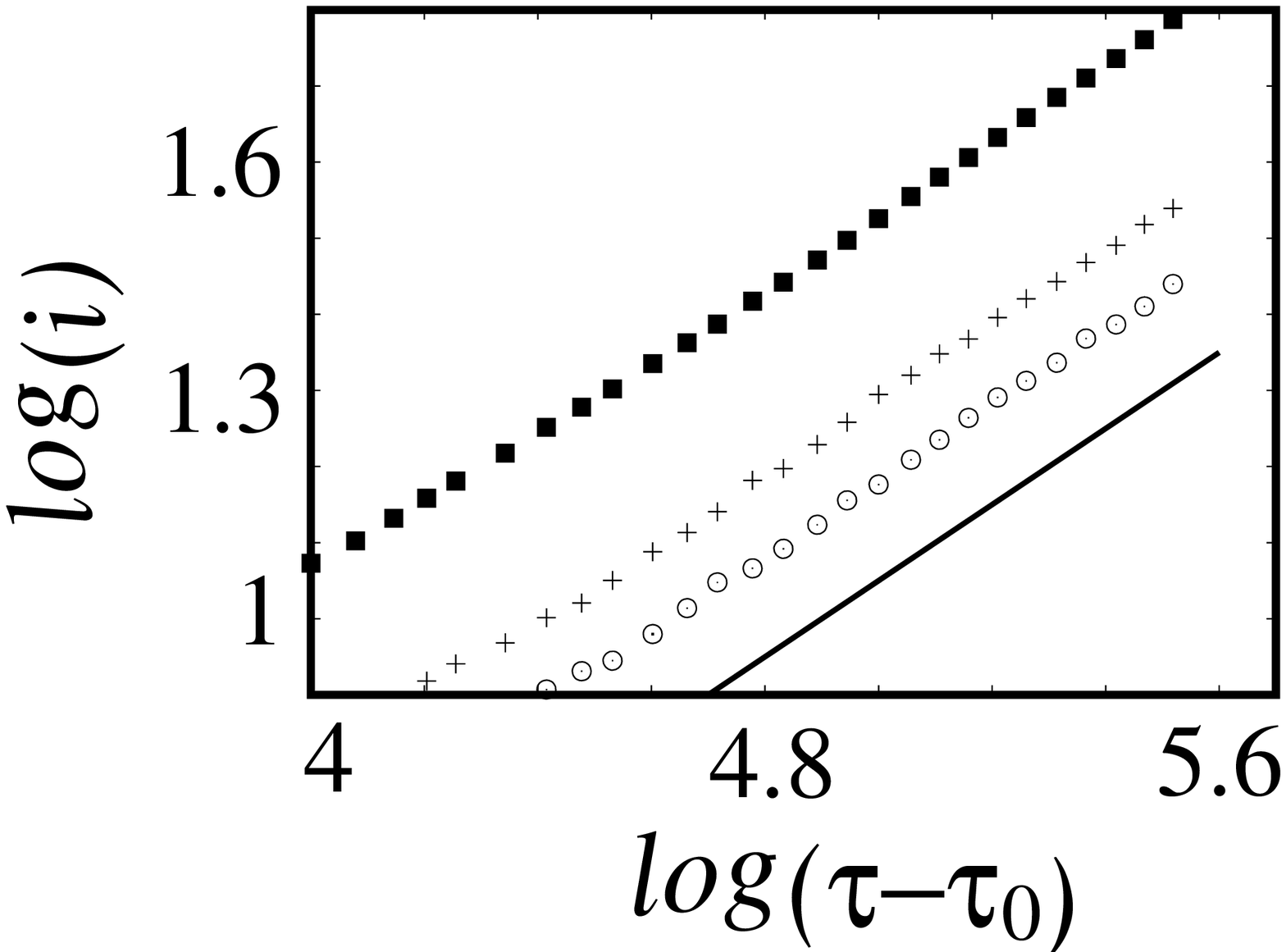 
}}{\vskip-0.0in}{ FIG. $6$~
The displacement of $x_1(\tau)$ after folding of the filament (Filled squares). 
Crosses shows the position of the fold along the contour. 
Open circles represent the arc length up to which $g^f$ is  linear. 
$(\tau-\tau_0)^{1/2}$  line (solid line) is given as a guide to the eye}\\

As the end moves, the position of the fold moves along the contour 
( see figure~$1$). Hence the fold length and the point at 
which tension falls to zero too follow  similar behavior as that 
of $x(s=0,\tau)$. 

To conclude, we have studied the dynamics of a single semiflexible filament under
the action of a compressing force at one end which lead to folding.
We have shown that the folding length($l_f$) of the filament 
is proportional to  $\sqrt \kappa$ and  decreases monotonically with  
the applied force $f$.  This follows from the fact that
the tension  in semiflexible filaments decay exponentially along  the contour.
It is shown that buckling  of a homogeneous elastic rod with  an exponential 
tension profile is qualitatively similar  to that of the  folding of a 
 semiflexible filament. 
The folding length  for a given $\kappa$ and force $\bf{f}$ does depend 
on the strength of the noise $\eta$. Thus once the initial configuration is 
chosen, a "zero temperature" calculation should give the same results.
As the filament folds the tension changes sign and we move into the 
stretching regime. We have shown, using numerical simulations and scaling arguments, 
 that in this regime the fold length, the end point $x_1(\tau)$ and the distance
along the contour to which tension has propagated, scales as $\tau^{1/2}$.

\noindent
 {\bf Acknowledgments}\\
We would like to thank S. K Srivatsa and Madan Rao for many discussions and 
comments. P.B.S thanks U.~Seifert for discussions during  a stay at 
MPIKG, Golm. We thank CSIR, India for financial support.\\


\noindent
\begin{thebibliography}{99}

\bibitem{strick} T. R. Strick {\it et al.}, Science, {\bf 271}, 1835 (1996)

\bibitem{ashkin} A. Ashkin, Science, {\bf 210}, 1081 (1980)

\bibitem{Dogte} M.\ Dogterom and B.\ Yurke,  Science {\bf 278}, 856 (1997) 

\bibitem{poptic}M.\ S.\ V.\ Kellermayer {\em et al.}, Science {\bf 276}, 1112 (1997); L.\ Tskhovrebova, {\em et al.}, Nature, {\bf 387}, 308 (1997)
 
\bibitem{doptic} S.\ B.\ Smith, Y.\ Cui, and C.\ Bustamante, Science
  {\bf 271}, 795 (1996); P.\ Cluzel {\em et al.}, Science
  {\bf 271}, 792 (1996)                                               

\bibitem{Maggs} A.C. Maggs,  Phys. Rev. E., {\bf 55}, 7396 (1996).

\bibitem{Frey} E. Frey, {\it Advances in Solid State Physics}, {\bf 41}, 345 (2001).

\bibitem{Alberts} See, e.g., B. Alberts {\it et al., Molecular Biology 
 of the Cell} (Garland Publ., New York, 1983).

\bibitem{Janmey} P. A. Janmey Curr. Op. Cell. Biol., {\bf 2}, 4 (1991), {\it Hand book 
of Biological Physics, Chapter 17, Page 805}, North Holland, Amsterdam (1995)

\bibitem{Gittes1} F. Gittes {\em et al}, J. Cell. Biol., {\bf 120}, 923 (1993)

\bibitem{Ott} A. Ott, M. Magnasco, A. Simon, and A. Libchaber, Phys. Rev. E {\bf 48}, R 1642 (1993)


\bibitem{landau}L. D. Landau and E. M. Lifshitz {\it Theory of Elasticity}
(Pergamon, Oxford, 1986) 

\bibitem{golu} L. Golubovi{\'c}, D. Moldovan and A. Peredera, Phys. rev. Lett. {\bf 81}, 3387 (1998)

\bibitem{Gittes2} F. Gittes {\em et al.},  Biophys. J. {\bf 70}, 418 (1996)

\bibitem{odijk} T. Odijk {\it J. Chem. Phys.}, {\bf 108}, 6923 (1998)     

\bibitem{ranjith} P. Ranjith and P. B. Sunil Kumar, {\it to be published}

\bibitem{frey} E. Frey, K. Kroy, J. Wilhelm and E. Sackmann. 
{\em  Dynamical Networks in Physics and Biology}, edited by D. Beysens and G.
    Forgacs (EDP Sciences - Springer, Berlin, 1998). 

\bibitem{Wang} M. D. Wang {\em et al}, Biophys. J., {\bf 72}, 1335 (1997)

\bibitem{bensimon} D.Bensimon, D. Dohmi, M. M{\'e}zard {\it Europhys. Lett.}, {\bf 42}(1), 97-102 (1998)     

\bibitem{hinge}P. Grassia and E. J. Hinch, J. Fluid Mech.  {\bf 308}, 255 (1996) 

\bibitem{Doi} M. Doi and S.F Edwards, { \it The Theory of Polymer Dynamics}
(Claredon Press, Oxford, 1986)

\bibitem{udo} U. Seifert, W. Wintz, and P. Nelson, Phys.Rev.Lett. {\bf 77},5389 (1996).  


\end{thebibliography}
\end{document}